\documentclass[aip,jap,amssymb,12pt]{revtex4-1}
\usepackage{graphicx}
\usepackage{epstopdf}
\usepackage[fleqn]{amsmath}
\usepackage{natbib}
\usepackage{endnotes}

\begin{document}

\title{Control of the magnetic in-plane anisotropy in off-stoichiometric NiMnSb}

\author{F. Gerhard, C. Schumacher, C. Gould, L.W. Molenkamp}
 \affiliation{Physikalisches Institut (EP3), Universit\"at W\"urzburg,\\ Am Hubland, D-97074 W\"urzburg, Germany}
\date{\today}

\begin{abstract}
NiMnSb is a ferromagnetic half-metal which, because of its rich anisotropy and very low Gilbert damping, is a promising candidate for applications in information technologies. We have investigated the in-plane anisotropy properties of thin, MBE-grown NiMnSb films as a function of their Mn concentration. Using ferromagnetic resonance (FMR) to determine the uniaxial and four-fold anisotropy fields, $\frac{2K_U}{M_s}$ and $\frac{2K_1}{M_s}$, we find that a variation in composition can change the strength of the four-fold anisotropy by more than an order of magnitude and cause a complete 90$^{\circ}$ rotation of the uniaxial anisotropy. This provides valuable flexibility in designing new device geometries.
\end{abstract}

\maketitle

\section*{Introduction}
NiMnSb is a half-metallic ferromagnetic material offering 100\% spin polarization in its bulk \cite{deGroot}, and was therefore long considered a very promising material for spintronic applications such as spin injection. Experience has shown however that preserving sufficiently high translation symmetry to maintain this perfect polarization at surfaces and interface is a major practical challenge, reducing its atractiveness for spin injection. The material nevertheless continues to be very promising for use in other spintronic applications; in particular in spin torque devices such as spin-transfer-torque (STT) controlled spin valves and spin torque oscillators (STO). This promise is based on its very low Gilbert damping, of order $10^{-3}$ or lower \cite{DissARiegler} which should enhance device efficiency, as well as on its rich and strong magnetic anisotropy which allows for great flexibility in device engineering. \newline
For example, it has been shown that STO oscillators formed from two layers of orthogonal anisotropy can yield significantly higher signal than those with co-linear magnetic easy axis \cite{Crozat, Dieny, Azzerboni, [{  }] [{ and references therein}]Mohseni}. Being able to tune the magnetic anisotropy of individual layers is clearly useful for the production of such devices. \newline
Previous results have shown a dependence of the anisotropy of NiMnSb on film thickness \cite{koveshnikov}, which offers some control possibilities when device geometries allow for appropriate layer thicknesses, but that is not always possible due to other design or lithography limitations. Here we show how the anisotropy of layers of a given range of thickness can effectively be tuned by slight changes in layer composition, achieved by adjusting the Mn flux. 

\section*{Experimental}
The NiMnSb layers are grown epitaxial by molecular beam epitaxy (MBE) on top of a 200 nm thick (In,Ga)As buffer on InP (001) substrates. All samples have a protective non-magnetic metal cap (Ru or Cu) deposited by magnetron sputtering before the sample is taken out of the UHV environment, in order to avoid oxidation and/or relaxation of the NiMnSb \cite{CKumpf}. The flux ratio Mn/Ni, and thus the composition, is varied between samples by adjusting the Mn cell temperature while the flux ratio Ni/Sb is kept constant. The thickness of most of the studied NiMnSb layers is \mbox{38$\pm$2} nm. Two samples have a slightly larger film thickness (45 nm, marked with ( ) in Fig. \ref{MvsLattice}a), caused by the change in growth rate due to the change in Mn flux. We verified that there is no correlation between anisotropy and sample thickness in this range.\newline
High Resolution X-Ray Diffraction (HRXRD) measurements of the (002) Bragg reflection are used to determine the vertical lattice constant of each sample. 
\begin{figure}
	\centering
		\includegraphics{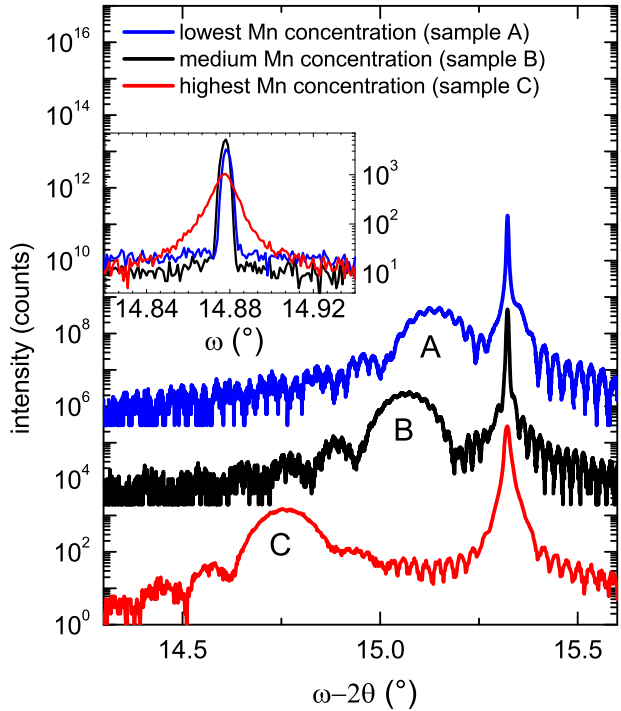}
		\caption{HRXRD $\omega$-$2\theta$-scans of 3 NiMnSb samples with various Mn concentrations. The curves are vertically offset for clarity. Inlet: $\omega$-scans showing high crystal quality.}
	\label{HRXRD}
\end{figure}
Fig.\ref{HRXRD} shows standard $\omega$-$2\theta$-scans of the (002) Bragg reflection on layers with the lowest and highest Mn concentrations used in the study, as well as a scan for a sample with medium Mn concentration. The sample with the lowest Mn content has a vertical lattice constant of \mbox{5.939 \r{A}} (sample A) and that with the highest Mn content (sample C) has a vertical lattice constant of \mbox{6.092 \r{A}}. To get an estimate of the vertical lattice constant of stoichiometric NiMnSb in our layer stacks, we used an XRD measurement of a stoichiometric, relaxed sample \cite{vRoy}. We determine a relaxed lattice constant of \mbox{$a_{rel}$ = (5.926 $\pm$ 0.007) \r{A}}. Together with the lattice constant of our InP/(In,Ga)As substrate, \mbox{5.8688 \r{A}}, and an estimated Poisson ratio of \mbox{0.3 $\pm$ 0.03}, we get the minimal and maximal values for the vertical lattice constant of stoichiometric NiMnSb: \mbox{$a_{\perp,max}$ = 5.999 \r{A}}, \mbox{$a_{\perp,min}$ = 5.957 \r{A}}. The vertical lattice constant of the sample with medium Mn concentration (sample B, \mbox{5.968 \r{A}}) lies in this range. We conclude that the composition of sample B is approximately stoichiometric. \newline 
In Ref. \onlinecite{Alling} and \onlinecite{Ekholm}, the effects of off-stoichiomteric defects in NiMnSb are discussed. Among the possible defects related to Mn, Mn$_{\text{Ni}}$ (Mn substituting Ni) is most likely (it has lowest formation energy) and the predicted decrease of the saturation magnetization is consistent with our observation (see Fig. \ref{MvsLattice}b). Furthermore, an increase of the lattice constant with increasing concentration of this kind of defect is predicted theoretically and observed experimentally. Thus, we can use the (vertical) lattice constant as a measure for the Mn concentration in our samples. \newline 
The crystal quality is also assessed by the HRXRD measurements. The inset in \mbox{Fig. \ref{HRXRD}} shows the $\omega$-scans of the same three NiMnSb layers. The $\omega$-scans of both the low and medium Mn concentration sample are extremely narrow with a full width half-maximum (FWHM) of 15 and 14 arcsec, respectively. A broadening for the sample with highest Mn concentration can be seen (FWHM of 35 arcsec). Reasons for the broadening can be partial relaxation of the layer due to the increased lattice mismatch with the (In,Ga)As Buffer, and/or defects related to the surplus of Mn. \newline
Using the experimental data of the lattice constant in Ref. \onlinecite{Ekholm}, we can estimate a difference in Mn concentration between sample A and C (extreme samples) of about 40\%. For sample C (extreme high Mn concentration), we determine a saturation magnetization of 3.4 $\mu_{Bohr}$ (see Fig. \ref{MvsLattice}b). According to Ref. \onlinecite{Ekholm}, this corresponds to a crystal where about 20\% of Ni is replaced by Mn. It should be noted that we investigated the effect of extreme surplus/deficit of Mn within the limits of acceptable crystal quality. As can be seen in Fig. \ref{MvsLattice}a, already a much smaller change in composition can change the strength and orientation of the magnetic anisotropy significantly. \newline
\begin{figure}
	\centering
		\includegraphics{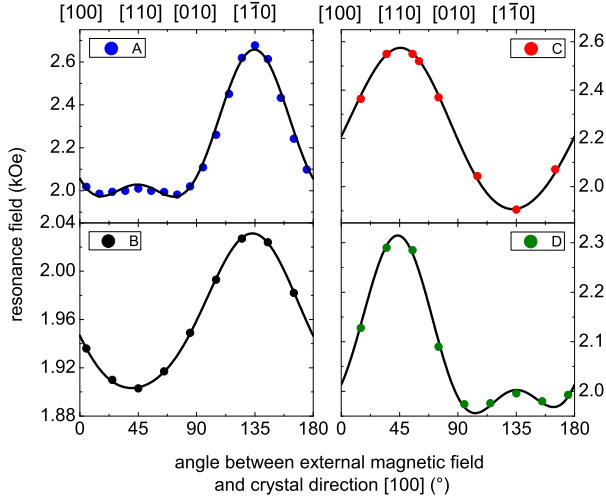}
		\caption{FMR measurements and simulation for four different samples. The symbols are measurements of the resonance frequency for magnetic fields along specific crystal directions, where 0$^{\circ}$ lies along [100]. The lines are simulations (see below) and also serve as a guide to the eye. Sample A, B and C correspond to the samples with lowest, medium and highest Mn concentration shown in Fig. \ref{HRXRD}. Sample D completes the various kinds of anisotropy observed in NiMnSb.}
	\label{Fitting}
\end{figure}
To map out the in-plane anisotropy of our samples, we use frequency-domain ferromagnetic resonance (FMR) measurements at a frequency of 12.5 GHz. The resonance fields are determined as a function of an external magnetic field applied at fixed angles ranging from 0$^{\circ}$ (defined as the [100] crystal direction) to 180$^{\circ}$. \mbox{Fig. \ref{Fitting}} shows results of these measurements for four different samples with four distinct types of anisotropy: Sample A and D both exhibit large uniaxial anisotropies with an additional four-fold component, however of opposite sign. The hard axis of sample A is along the [$1\bar{1}0$] crystal direction, where for sample D the hard axis is along the [$110$] crystal direction. Sample B and C both show mainly uniaxial anisotropies, again with opposite signs. \newline
The FMR data can be simulated with a simple phenomenological magnetostatic model to extract the anisotropy components (derivation taken from Ref. \onlinecite{Heinrich}). The free energy equation for thin films of cubic materials is given by:
\begin{equation}
	\epsilon_c = - \frac{K_1^\parallel}{2}(\alpha_x^4+\alpha_y^4)-\frac{K_1^\perp}{2}\alpha_z^4-K_u\alpha_z^2
\label{freeenergy}
\end{equation}
where $\alpha_x$, $\alpha_y$ and $\alpha_z$ describe the magnetization with respect to the crystal directions [100], [010] and [001]. $K_1^\parallel$ is the four-fold in-plane anisotropy constant, $K_u$ and $K_1^\perp$ represent the perpendicular uniaxial anisotropy (second and fourth order respectively). In our in-plane FMR geometry, the fourth order perpendicular anisotropy term $K_1^\perp$ can be neglected. Instead, an additional uniaxial in-plane anisotropy term is added: 
\begin{equation}
	\epsilon_u = -K_u^\parallel\frac{(\hat{n}\cdot\hat{M})^2}{M_s^2} \nonumber\\
		\label{Kuniaxial}
\end{equation}
with the unit vector $\hat{n}$ along the uniaxial anisotropy and the saturation magnetization $M_s$, $\hat{M}$.
The Zeeman term coupling to the external field $H_0$ and a demagnetization term originating from the thinness of the sample, are defined as 
\begin{equation}
 \epsilon_{Z}=-\mathbf{M}\cdot\mathbf{H_0} \quad, \quad \epsilon
_{demag}= -\frac{4\pi D M_\perp^2}{2}  
\label{ZeemanDemag} 
\end{equation}
and added as well to the free energy. 
The effective magnetic field
\begin{equation}
	H_{eff}=-\frac{\partial \epsilon_{total}}{\partial \mathbf{M}}
	\label{Heff}
\end{equation}
with
\begin{equation}
\epsilon_{total} = \epsilon_c + \epsilon_u + \epsilon_Z + \epsilon_{demag} 
\end{equation}
is used to solve the Landau-Lifshitz-Gilbert-Equation (LLG):
\begin{equation}
-\frac{1}{\gamma}\frac{\partial M}{\partial t}=[M\times H_{eff}]-\frac{G}{\gamma^2M_s^2}[M\times\frac{\partial M}{\partial t}]
\label{LLG}
\end{equation}
with the gyromagnetic ratio $\gamma = \frac{g\mu_B}{\hbar}$ and the Gilbert damping constant $G$.
The resonance condition can be found by calculating the susceptibility\cite{Kittel}, $\chi=\frac{\partial M}{\partial H}$:
\begin{equation}
(\frac{\omega}{\gamma})^2=B_{eff}\mathcal H^*_{eff}
\label{resonance}
\end{equation}
In the following, we neglect the damping contribution since $\frac{G}{\gamma M_s}$ in our samples is of the order of $10^{-3}$ or lower. Thus, $B_{eff}$ and $\mathcal H^*_{eff}$ in our case can be found to be:
\begin{eqnarray}
&\mathcal H^*_{eff}=H_0 cos[\phi_M-\phi_H]+\frac{2K_1^{\parallel}}{M_s}cos[4(\phi_M-\phi_F)]\nonumber\\
&+\frac{2K_U^{\parallel}}{M_s}cos[2(\phi_M-\phi_U)] 
\end{eqnarray}
\begin{eqnarray}
&B_{eff}=H_0cos[\phi_M-\phi_H]+\frac{K_1^{\parallel}}{2M_s}(3+cos[4(\phi_M-\phi_F)])\nonumber\\
&+4\pi D M_s - \frac{2K_U^{\perp}}{M_s}+\frac{K_U^{\parallel}}{M_s}(1+cos[2(\phi_M-\phi_U])) 
\label{Beff}
\end{eqnarray}
Here, $\phi_M$, $\phi_H$ and $\phi_U$ define the angles of the magnetization, external magnetic field and in-plane easy axis of the uniaxial anisotropy, respectively, with respect to the crystal direction [100]. $\phi_F$ accounts for the angle of the four-fold anisotropy. At the magnetic fields used in these studies, it is safe to assume $\phi_M = \phi_H$ \cite{[{  }] [{We have confirmed from frequency dependent measurements that this assumption leads to errors smaller than the size of the symbols in Fig. \ref{MvsLattice}a.}]phi}. In equation (\ref{Beff}), $4\pi D M_s - \frac{2K_U^{\perp}}{M_S}$ can be defined as an effective magnetization $4\pi M_{eff}$, containing the out-of-plane anisotropy. It is used as a constant in our simulation. 
\newline
For each sample, we extract $\frac{2K_1}{M_s}$  and $\frac{2K_U}{M_s}$, the four-fold and uniaxial in-plane anisotropy field, from the simulation and plot them versus the vertical lattice constant (Fig. \ref{MvsLattice}a). The vertical, dotted lines mark the range where stoichiometric NiMnSb is expected. 
\begin{figure}[ht]
	\centering
		\includegraphics{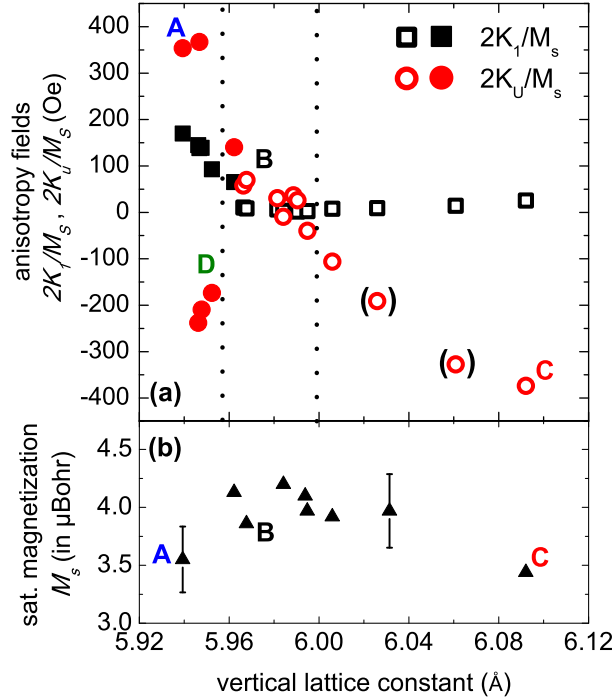}
		\caption{(a) Uniaxial anisotropy field $\frac{2K_U}{M_s}$ and four-fold anisotropy field $\frac{2K_1}{M_s}$ for NiMnSb layers with various Mn concentrations. The vertical lattice constant is used as a gauge of the Mn content. Samples with a rotated RHEED pattern (see last section) are indicated by open symbols. The dotted lines mark the range where stoichiometric NiMnSb is expected. The samples of lowest, medium and highest Mn concentration (A, B and C) together with sample D are marked. The two samples marked with ( ) exhibit slightly higher film thickness than the other samples. (b) Saturation magnetization $M_s$ depending on the vertical lattice constant. }
	\label{MvsLattice}
\end{figure}
For vertical lattice constants in the range from 5.96 to 6.00 \r{A}, both anisotropy fields are relatively small. The four-fold contribution increases for samples with decreasing vertical lattice constant (lower Mn concentration) but remains small for larger vertical lattice constant (increasing Mn concentration). The uniaxial anisotropy gets more strongly negative with increasing vertical lattice constant, whereas in samples with lower vertical lattice constants, the uniaxial field can be either positive or negative while its absolute value grows significantly with decreasing vertical lattice constant. The change in sign of the uniaxial anisotropy field at a vertical lattice constant of about 5.99 \r{A} corresponds to a rotation of the easy axis from the [$110$] direction (positive anisotropy fields) to the [$1\bar{1}$0] direction. One can see that already a small change of the vertical lattice constant (small change in composition) is sufficient to rotate the uniaxial anisotropy as well as to induce a significant four-fold anisotropy. \newline
The fitting accuracy of the extracted anisotropy fields is $\sim$5\%, giving error bars smaller than the symbols in \mbox{Fig. \ref{MvsLattice}a}. It should be noted that in order to exactly extract the anisotropy constants $K_1$ and $K_U$ from the anisotropy fields, the saturation magnetization $M_s$ of each sample is needed. This can be determined by SQUID measurements. We have performed such measurements on a representative fraction of the samples (Fig. \ref{MvsLattice}b). Samples with medium Mn concentration show saturation magnetizations which, to experimental accuracy of about 8\% are consistent with the theoretically expected 4.0 $\mu_{Bohr}$ per unit formula for stoichiometric NiMnSb \cite{Graf}. The estimated measurement accuracy of ~8\% accounts for uncertainty in the sample thickness extracted from the HRXRD data of about 5\%, as well as errors in determining the exact sample area, SQUID calibration and SQUID response due to finite sample size. Our samples with highest and lowest magnetization show a slight decrease in saturation magnetization, of order 12\%. This change is sufficiently small to be neglected in the overall assesment of the anisotropy vs. vertical lattice constant of Fig. \ref{MvsLattice}a.
\newline
In an attempt to understand the effect of higher or lower Mn concentration on the crystal structure in our samples, we consider the possible non-stoichiometric defects which can exist in NiMnSb, as discussed in Ref. \onlinecite{Alling}. Formation energies, magnetic moment change and effect on the half-metallic character are presented there for each type of defect. Mn-related defects are a) Mn substituting Ni or Sb (Mn$_{\text{Ni}}$, Mn$_{\text{Sb}}$), b) Mn on a vacancy position (Mn$_{\text{I}}$), c) Ni or Sb substituting Mn (Ni$_{\text{Mn}}$, Sb$_{\text{Mn}}$) or d) a vacancy position at the Mn site (vac$_{\text{Mn}}$). With a surplus of Mn, both Mn substituting Ni or Sb and Mn incorporated on the vacancy position seem plausible. However, the formation energy of Mn$_{\text{Sb}}$ is more than three times larger than for the other defects, suggesting it should be very rare. On the other hand, in the case of a Mn deficiency, either Ni or Sb could substitute Mn or vacancies can be built into the crystal. All those three defects have similar formation energies, making them equally possible. 
\newline
Except for Mn$_{\text{I}}$ and Mn$_{\text{Sb}}$, all of these possible defects reduce the magnetic moment per formula unit. Our observations of a lower magnetic moment for samples with either high or low Mn flux, are thus consistent with the defects Mn$_{\text{Ni}}$, Ni$_{\text{Mn}}$, Sb$_{\text{Mn}}$ and vac$_{\text{Mn}}$. The positive contribution of Mn$_{\text{I}}$ to the magnetic moment is however some 5 times smaller than the decrease induced by the other defects, so some fraction of defects of the Mn$_\text{I}$ variety could also be present in the samples. A detailed discussion on the transition from stoichiometric NiMnSb towards off-stoichiometric Ni$_\text{1-x}$Mn$_\text{1+x}$Sb is given in Ref. \onlinecite{Ekholm}. It is shown that the lattice constant of off-stoichiometric NiMnSb increases for increasing substitution of Ni by Mn. This behavior is clearly seen in our samples for increasing Mn concentration and we conclude that this kind of defect is most prominent in our samples. An explanation for a decreasing lattice constant for decreasing Mn concentration is yet to be found. 
\newline
A further observation which may provide insight into the observed anisotropy behavior comes from Reflective High Energy Electron Diffraction (RHEED), which is used to monitor the surface of the sample in-situ during the growth. 
\begin{figure}
	\centering
		\includegraphics{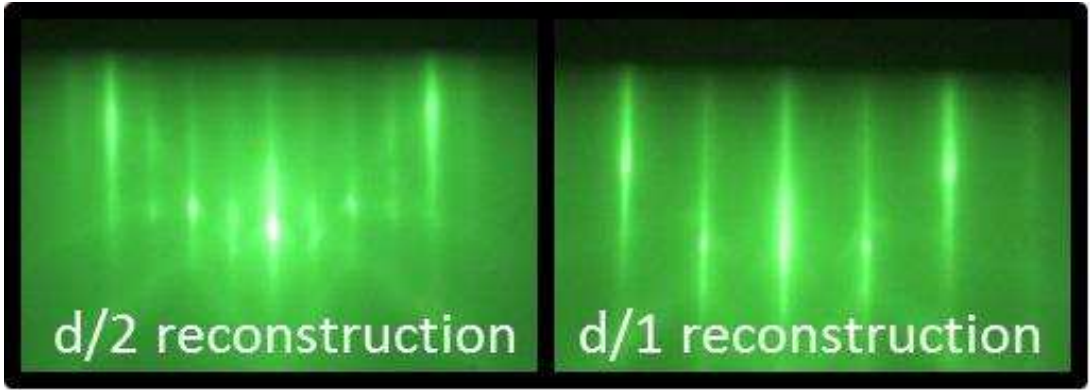}
		\caption{Typical RHEED reconstruction of the NiMnSb surface illustrating the two reconstructions discussed in the text.}
		\label{RHEED}
\end{figure}
RHEED provides information about the surface reconstruction, which turns out to be sensitive to the Mn content. In all samples, at the beginning of the growth (after approximately one minute), the surface reconstruction exhibits a clear $2\times1$ pattern, meaning a $d/2$ reconstruction in the [110] crystal direction and a $d/1$ reconstruction along [$1\bar{1}0$] direction (see Fig.\ref{RHEED}). How this pattern then evolves during growth depends on the Mn flux. For ideal Mn flux, the pattern is stable throughout the entire 2 hour growth time corresponding to a 40 nm layer.  A reduced Mn flux results in a more blurry RHEED pattern, but does not lead to any change in the surface reconstruction. A higher Mn flux, on the other hand, causes a change of the reconstruction such that the d/2 pattern also becomes visible along the [$1\bar{1}0$] direction and fades over time in the [$110$] direction until a 90$^{\circ}$ rotation of the original pattern has been completed. The length of time (and thus the thickness) required for this rotation depends strongly on the Mn flux. A slightly enhanced Mn flux causes a very slow rotation of the reconstruction that can last the entire growth time, whereas a significant increase of the Mn flux (sample with vertical lattice constants above 6.05 \AA) will cause a rotation of the reconstruction within a few minutes of growth start, corresponding to a thickness of only very few monolayers. Based on these observations, our samples can be split into two categories: samples with a stable $2\times1$ reconstruction and those with a $2\times1$ reconstruction that rotates during growth. In Fig. 3a, samples with a stable RHEED pattern are indicated with filled symbols while empty symbols show samples with a rotated RHEED reconstruction. It is interesting to note that all samples with a rotated reconstruction exhibit a very low four-fold anisotropy field. In addition, the sooner the rotation of the RHEED pattern occurs, the stronger the uniaxial anisotropy is.

\section*{Summary}
We have shown that the anisotropy of NiMnSb strongly depends on the composition of the material. A variation of the Mn flux results in different (vertical) lattice constants (measured by HRXRD) that can be used for a measure of the Mn concentration. RHEED observations (in-situ) during the growth already give an indication of high or low Mn concentration. The anisotropy shows a clear trend for increasing Mn content. Using this together with the RHEED observations, NiMnSb layers with high crystal quality and anisotropies as-requested can be grown. The microscopic origin of this behavior remains to be understood, and it is hoped that this paper will stimulate further efforts in this direction. The phenomenology itself is nevertheless of practical significance in that it provides interesting design opportunities for devices such as spin-valves that could be made of two NiMnSb layers with mutually parallel or orthogonal magnetic easy axes as desired. 

\section*{ACKNOWLEDGEMENTS}
We thank T. Naydenova for assistance with the SQUID measurements. This work was supported by the European Commission FP7 contract ICT-257159 ``MACALO''.


%

\end{document}